\documentclass[fleqn,12pt,twoside]{article}
\usepackage[headings]{espcrc1}
\readRCS
$Id: espcrc1.tex,v 1.2 2004/02/24 11:22:11 spepping Exp $
\usepackage{graphicx}
\usepackage[figuresright]{rotating}

\newcommand{ \be }{\begin{equation}}
\newcommand{ \ee }{\end{equation}}
\newcommand{ \bea }{\begin{eqnarray}}
\newcommand{ \eea }{\end{eqnarray}}
\newcommand{ \la }{\langle}
\newcommand{ \ra }{\rangle}

\newcommand{ \dpti }{\delta p_{t,1}}
\newcommand{ \dptj }{\delta p_{t,2}}
\newcommand{ \bp }{{\bf p}}
\newcommand{ \mpt }{{\la p_t \ra}}

\usepackage{color}

\newcommand{\AmS}{{\protect\the\textfont2
  A\kern-.1667em\lower.5ex\hbox{M}\kern-.125emS}}

\title{Two particle rapidity, transverse momentum, 
and azimuthal correlations in relativistic
nuclear collisions and transverse radial expansion}

\author{Sergei A.~Voloshin\address[WSU]{Department of Physics and Astronomy, 
        Wayne State University, \\ 
        666 W. Hancock, Detroit, Michigan 48201}%
        }
       
\runtitle{Two particle correlations in nuclear collisions}
\runauthor{S. Voloshin}

\begin{document}

\maketitle

\begin{abstract}
At the very first stage of an ultra-relativistic nucleus-nucleus collision
new particles are produced in individual nucleon-nucleon collisions. 
In the transverse plane, all particles 
from a single $NN$ collision are initially located at the same position. 
The subsequent transverse radial expansion of the system creates
strong position-momentum correlations and
leads to characteristic rapidity, transverse momentum, and
azimuthal correlations among the produced particles. 
\end{abstract}

\section{Introduction}
The physics of the high energy heavy ion collisions attracts strong
attention of the physics community 
as creation of a new type of matter, the Quark-Gluon Plasma,
is expected in such collisions. 
During the last few years of Au+Au collisions at the BNL Relativistic
Heavy Ion Collider (RHIC) many new phenomena has been observed,
such as strong elliptic flow~\cite{star-flow1} and  
suppression of the high transverse momentum two particle back-to-back 
correlations~\cite{star-highpt}.
These observations strongly indicate that a dense partonic matter 
has been created in such high energy nuclear collisions.
Parton reinteractions lead to pressure build-up and
the system undergoes longitudinal and transverse expansion,
the latter leading to an increase in
the particle final transverse momenta.
Usually the transverse expansion is studied via detailed
analysis of single particle transverse momentum spectra, 
most often using thermal
parameterization suggested in~\cite{thermal-ssh}.
In this paper we note that the transverse radial expansion should also lead
to characteristic rapidity, transverse momentum, and azimuthal angle
two-particle correlations.

At the first stage of a AA collision many individual 
nucleon-nucleon collision happen. 
The subsequent transverse expansion of the system 
creates strong position-momentum correlations 
in the transverse plane: further from 
the center axis of the system a particle is produced initially,
on average the larger push it gets from 
other particles during the system evolution. 
The 'push' is in the transverse direction, and on average does not affects
the longitudinal momentum component.
As all particles produced in
the same $NN$ collision have initially 
the {\em same spatial} position in the transverse plane, 
they get {\em on average} the
same push and thus become correlated.
As discussed below this picture leads to many distinctive phenomena, most 
of which can be studied by means of two (and many-) particle
correlations. The correlations can extend over wide rapidity range,
but are not boost invariant, as transverse expansion itself depends
on rapidity.

\section{Two particle transverse momentum correlations}
The single particle spectra are affected by radial flow in such a way
that the mean transverse momentum
is  mostly sensitive to the {\em average} expansion velocity squared
$
\la p_t\ra_{AA} \approx \la p_t \ra_{NN} + \alpha \la v^2 \ra,
$
(see Fig.~1 below) and to
much lesser extend to the actual velocity profile (dependence of the
expansion velocity on the radial distance from the center axis 
of the system).    
The two-particle transverse momentum correlations~\cite{vkr}, 
$\la p_{t,1} p_{t,2} \ra - \la  p_{t,1}  \ra \la p_{t,2} \ra
\equiv \la \delta p_{t,1} \, \delta p_{t,2} \ra$, would be
sensitive mostly to the {\em variance}
 in collective transverse expansion velocity, and thus are more
sensitive to the actual velocity profile:
\be
\la \delta p_{t,1} \delta p_{t,2} \ra_{AA}  \approx
D_{N_{c}}
(\la \delta p_{t,1} \delta p_{t,2} \ra_{NN} + \alpha^2
\sigma^2_{v^2});
\;D_{N_{c}} =
\frac{\la n(n-1) \ra_{NN}}
{(N_{c}-1) \la n \ra_{NN}^2 + \la n(n-1) \ra_{NN}}.
\label{edpt}
\ee 
The factor $D$
takes into account the dilution of the
correlations due to a mixture of particles from $N_{c}$ uncorrelated 
$NN$ collisions, and
that in an individual $NN$ collision the mean number of particle pairs,
$\la n(n-1) \ra_{NN}$, on average is larger than 
$ \la n \ra_{NN}^2$; at ISR energies in central rapidity region
$\la n(n-1) \ra_{NN} \approx 1.66 \la n \ra_{NN}^2$~\cite{Foa}.   

We employ a thermal model~\cite{thermal-ssh} for further calculations. 
In this model particles are produced by
freeze-out of the thermalized matter at temperature $T$, approximated 
by a boosted Boltzmann distribution. 
Assuming boost-invariant longitudinal expansion and freeze-out 
at constant proper time, one finds
\be
\frac{dn}{d\bp_t} \sim 
\int d\rho_t d\phi_b \rho_t^{2/n-1} J(\bp_{t};T,\rho_t,\phi_b);\;\;
 J(\bp_{t};T,\rho_t,\phi_b) \equiv m_t 
K_1(\beta_t) e^{\alpha_t \cos(\phi_b-\phi)},
\ee
where $\rho_t$ is the transverse flow rapidity, 
$\phi_b$ is the boost direction,
$\alpha_t=(p_t/T)\sinh(\rho_t)$, and $\beta_t=(m_t/T)\cosh(\rho_t)$.
It also assumes a uniform matter density within a cylinder, $r<R$, and a
power law transverse rapidity flow profile $\rho_t \propto r^n$.
Two particle spectrum for particles originating from the same $NN$
collision
would be
\be
\frac{dn_{pair}}{d\bp_{t,1} d\bp_{t,2}} \sim 
\int d\rho_t d\phi_b \rho_t^{2/n-1} 
J(\bp_{t,1};T,\rho_t,\phi_b)
J(\bp_{t,2};T,\rho_t,\phi_b)
 .
\ee
It additionally assumes that during the expansion time (before the
freeze-out) the particles produced originally at the same spatial position
 do not diffuse far one from another compared to the system size.

The results of the numerical calculations based on the above equations
are presented in Fig.~1 as function of 
$\la \rho_t^2 \ra = \la \rho_t\ra^2 (4n+4)/(2+n)^2$. 
The results are shown for two different
velocity (transverse rapidity) profiles, $n=$2, and $n=0.5$.  
\begin{figure}
  \includegraphics[width=0.48\textwidth]{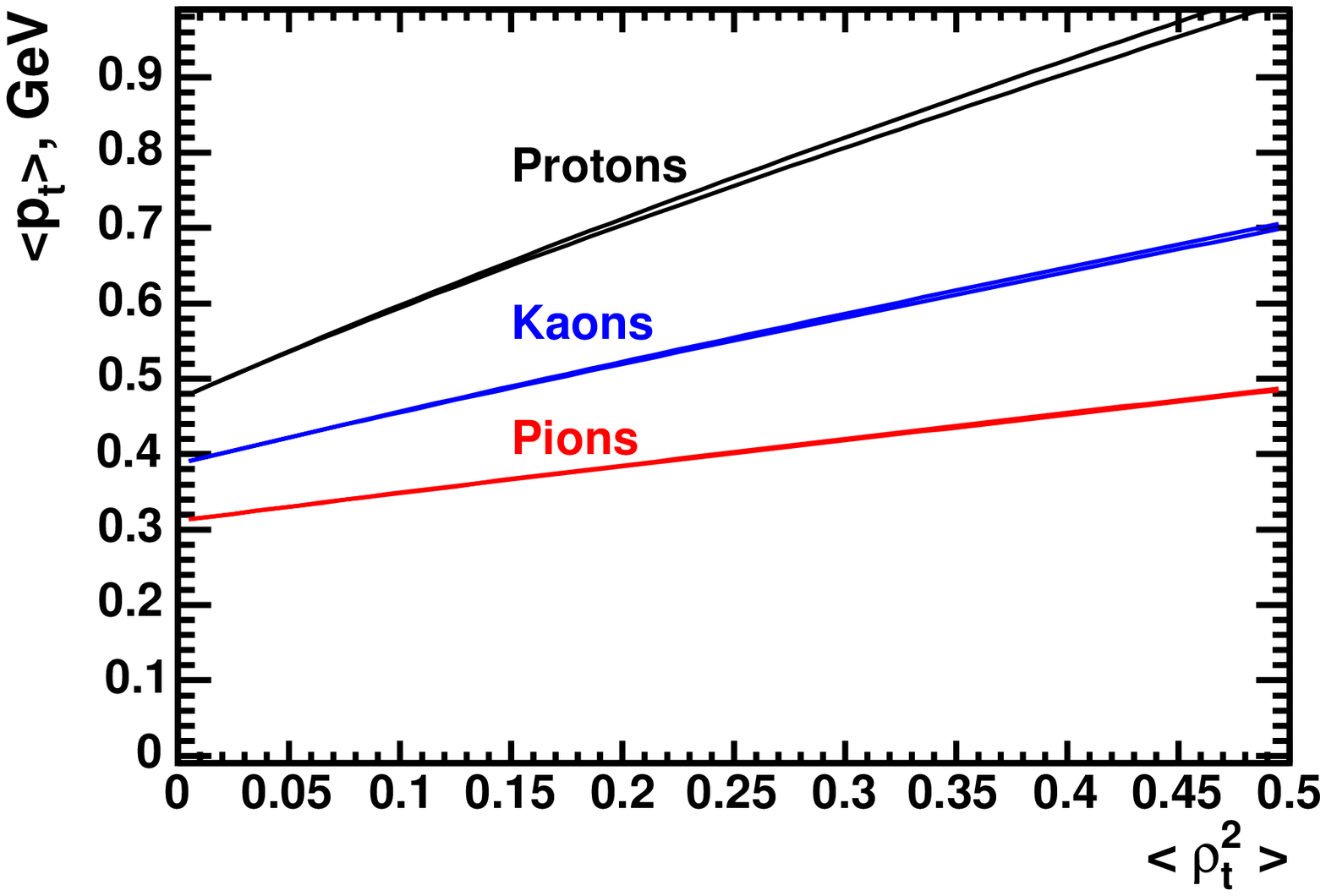}
  \includegraphics[width=0.48\textwidth]{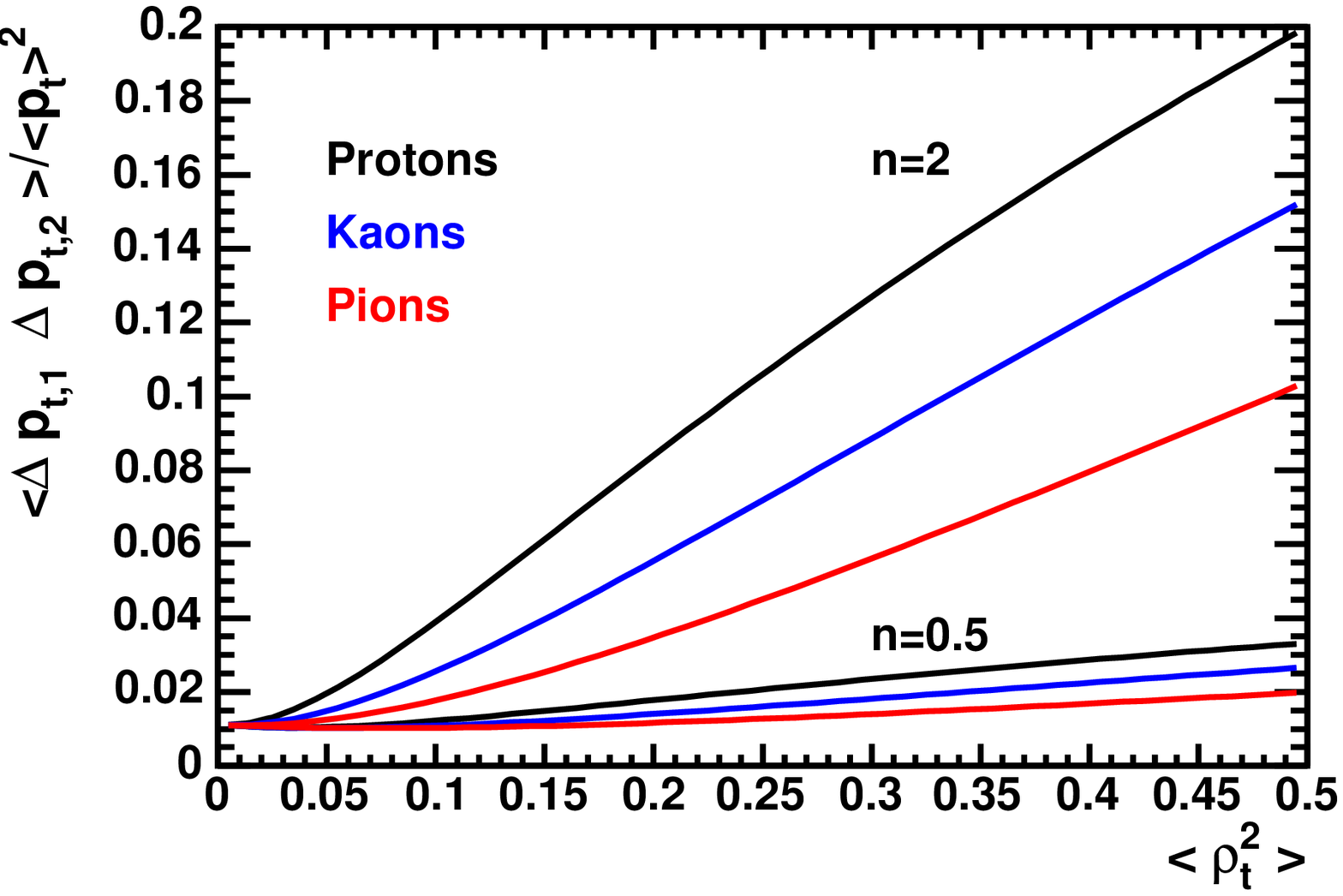}
	\vspace*{-11mm}
  \caption{
(color online) Mean transverse momentum and two particle $p_t$
  correlations in the blast wave calculations. $T=110$~MeV.}
  \label{fig1}
\end{figure}
One observes that indeed for all the particle types presented, $\mpt$
depends very
weakly on the actual profile. On opposite, the correlations
are drastically different for two cases studied.

In Fig.~2 we compare our estimates with preliminary STAR data~\cite{gary04} on
two particle $p_t$ correlations. 
We use the mean expansion velocity and temperature parameters
from~\cite{star-spectra} and assume $N_c=N_{part}/2$
and $\la \dpti \dptj \ra /\la p_t\ra^2 =0.011$ (about 20\% smaller
than measured at ISR~\cite{isrpp}. It is observed that the transverse
expansion with linear velocity profile produces too strong correlations.
\begin{figure}[htb]
\begin{minipage}[t]{78mm}
  \includegraphics[width=0.99\textwidth]{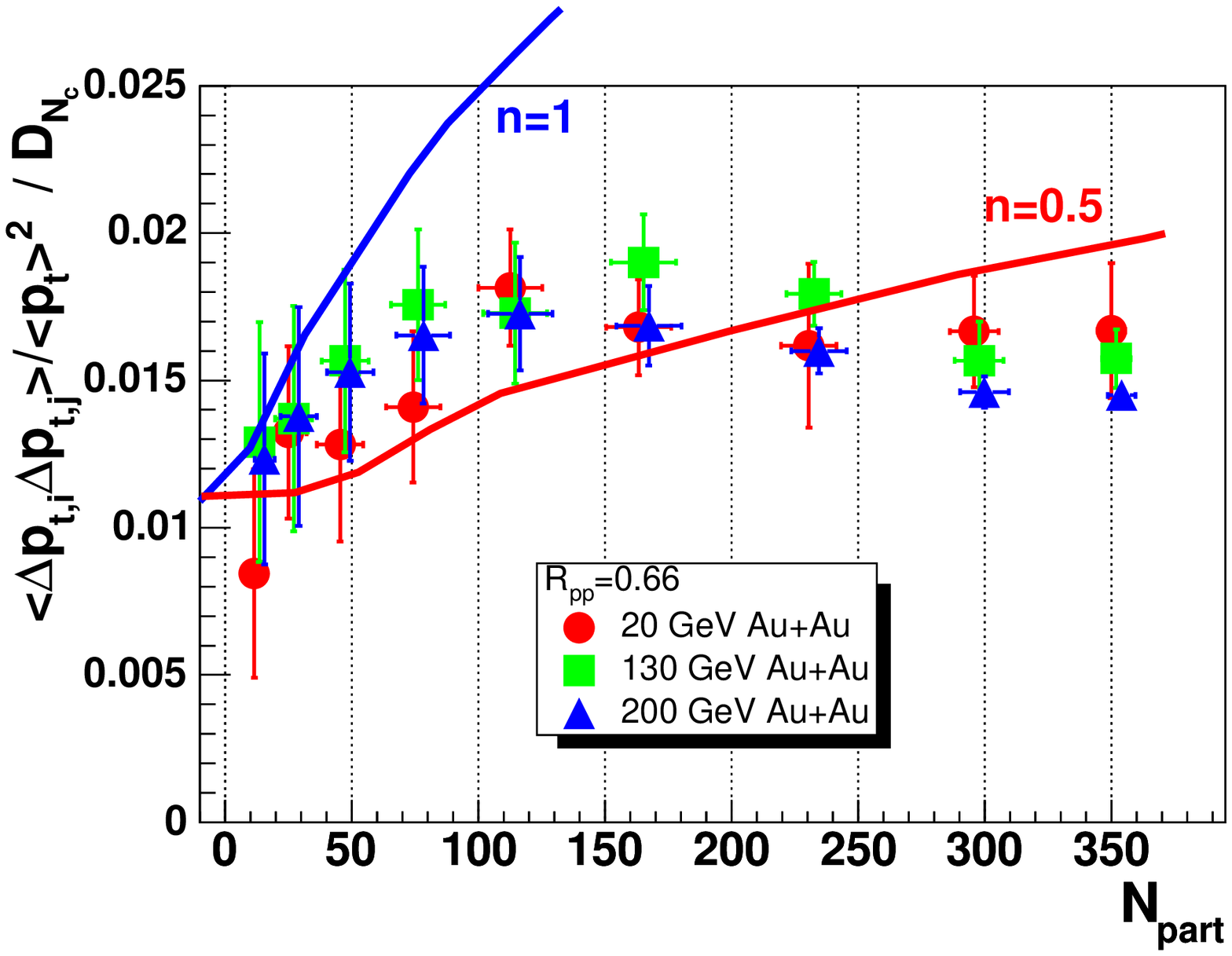}
	\vspace*{-14mm}
  \caption{
Comparison of Blast Wave calculations for two different velocity profiles 
with preliminary STAR data~\cite{gary04}.  Relation
$\la n(n-1)\ra_{NN} =1.66 \la n\ra^2_{NN}$ has been used.
} 
  \label{fig3}
\end{minipage}
\hspace{\fill}
\begin{minipage}[t]{77mm}
  \includegraphics[width=0.9\textwidth]{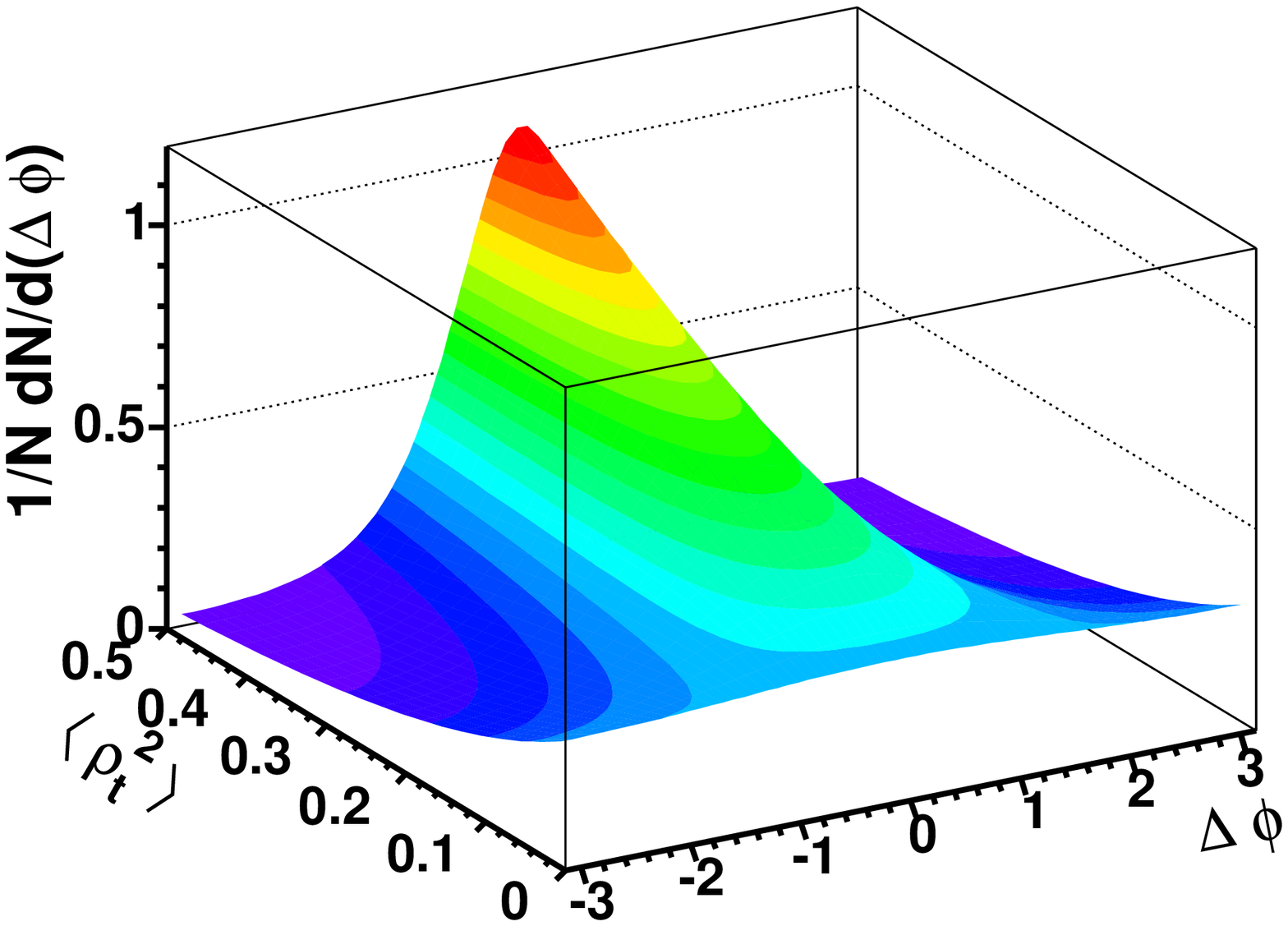}
	\vspace*{-9mm}
  \caption{(color online) Two pion $\Delta \phi$ distribution as
  function of $\la \rho_t^2 \ra$ in the Blast Wave model. 
  Linear velocity profile and
  $T=110$~MeV have been assumed.}
  \label{fig2}
\end{minipage}
\end{figure}

\section{Rapidity and azimuthal correlations. Charge balance functions.}
The rescatterings during transverse expansion lead not only to 
the increase of the transverse momentum but also 
to the particle diffusion in the rapidity space. 
As we neglect such a diffusion in the current analysis, 
the narrowing of the charge balance 
function~\cite{balance} observed in~\cite{star-balance1} 
would be described just by increase in mean $p_t$ as~\cite{balance2}:
\be
\Delta p_z=m_t \sinh(\Delta y) \approx m_t \Delta y \approx const.
\ee
This effect is consistent with experimentally observed
narrowing for about 15 -- 20\% of the balance function width 
with centrality~\cite{star-balance1} and with
centrality dependence of the net charge fluctuations~\cite{star-mult-fluc}.

As all particles from the same $NN$ collisions are pushed in the same
direction (radially in the transverse plane) they become correlated 
in azimuthal space. 
The correlations can become really strong for
large transverse flow as
shown in Fig.~2 (again, for particles originated from
the {\em same} $NN$ collision).  
The second harmonic in the azimuthal correlations generated by radial 
expansion is of a particular interest as it would contribute 
to the measurements of elliptic flow. 
The numbers from Fig.~3 corresponding to $\la \rho_t^2 \ra \sim0.3$ 
are comparable with the estimates 
of the strength of non-flow type azimuthal correlation 
estimates~\cite{star-flow-PRC}. 
Thus the azimuthal correlations generated by transverse expansion could 
be a major contributor to the non-flow azimuthal correlations.

The magnitude of the correlations due to transverse expansion should 
be sensitive to the system thermalization time and the particle
diffusion in the thermalized matter during the expansion; 
the azimuthal correlations would be the most interesting/useful
 for such a study.
Another application can be in ``jet tomography'' to infer information
about how ``deep'' in the system the hard collision has occurred.
For that, one has to correlate the jet (high $p_t$ hadron) yield
with mean transverse momentum of particles taken at different rapidity
(but better at similar azimuthal angle). 
In the same $NN$ collision where the high $p_t$ particle is emitted, the soft
particles are produced as well. Those
soft particles experience the transverse 'push' corresponding
to the spatial position in the transverse plane where 
the original $NN$ collision happens to be. 
Then the mean transverse momentum of the associated
particles would provide the information on how close to the center 
of the system the collision occurred.      

\vspace*{2mm}
The above described picture of $AA$ collisions has many
interesting observable effects, only a few mentioned in this paper. 
The picture become even richer if one looks at the 
identified particle correlations.
Many  questions require a
detailed model study, but the approach opens a potentially very interesting
possibility to address the initial conditions and the subsequent
evolution of the system created in an $AA$ collision.  

{\it Acknowledgments.}
Discussions with R.~Bellwied, S.~Gavin, C.~Pruneau, S. Pratt, and
U.~Heinz are gratefully acknowledged. 
This work was supported in part by the
U.S. Department of Energy Grant No. DE-FG02-92ER40713.

 \end{document}